# Clio-X: A Web3 Solution for Privacy-Preserving AI Access to Digital Archives

VICTORIA L. LEMIEUX, University of British Columbia,, , Canada
ROSA GIL, University of Lleida, , Spain
FAITH MOLOSIWA, University of British Columbia,, , Canada
QIHONG ZHOU, University of British Columbia,, , Canada
BINMING LI, University of British Columbia,, , Canada
ROBERTO GARCIA, University of Lleida,, , Spain
LUIS DE LA TORRE CUBILLO, UNED, Spain
ZEHUA WANG, University of British Columbia,, , Canada

As archives turn to artificial intelligence (AI) to manage growing volumes of digital records, privacy risks inherent in current AI data practices raise critical concerns about data sovereignty and ethical accountability. This paper explores how privacy-enhancing technologies (PETs) and Web3 architectures can support archives to preserve control over sensitive content while still being able to make it available for access by researchers. We present Clio-X, a decentralized, privacy-first Web3 digital solution designed to embed PETs into archival workflows and support AI-enabled reference and access. Drawing on a user evaluation of a medium-fidelity Clio-X prototype, the study reveals both interest in the potential of the solution and significant barriers to adoption related to trust, system opacity, economic concerns, and governance. Using Rogers' Diffusion of Innovations theory, we analyze the sociotechnical dimensions of these barriers and propose a path forward centered on participatory design and decentralized governance through a Clio-X Decentralized Autonomous Organization (DAO). By integrating technical safeguards with community-based oversight, Clio-X offers a novel model to ethically deploy AI in cultural heritage contexts.

Authors' Contact Information: Victoria L. Lemieux, v.lemieux@ubc.ca, University of British Columbia,, Vancouver,, Canada; Rosa Gil, University of Lleida,, Spain; Faith Molosiwa, University of British Columbia,, Vancouver,, Canada; Qihong Zhou, University of British Columbia,, Vancouver,, Canada; Binming Li, University of British Columbia,, Vancouver,, Canada; Roberto Garcia, University of Lleida,, Lleida,, Spain; Luis de la Torre Cubillo, UNED, Madrid, Spain; Zehua Wang, University of British Columbia,, Vancouver,, Canada.

## 1 Introduction

Just as in other sectors of society, artificial intelligence (AI) is being applied widely in the cultural sector, particularly by archives seeking efficient ways to administer large bodies of archival documentation in digital form [1], [2], [3]. Building upon work presented in [4], this paper seeks to demonstrate how archives and other cultural institutions can use privacy-enhancing technologies and practices to address concerns about AI data practices, focusing particularly on the issue of AI's potential for privacy violations and loss of data control.

As reported in [4], archival institutions hold a great volume of archival documents that remain sensitive and must be protected by law or custom. To protect such documents, archivists undertake 'sensitivity reviews' to identify and redact sensitive content. A mounting accumulation of documents, however, prevents archivists from undertaking such reviews in a timely manner, which reduces the accessibility of archival holdings leading to the formation of so-called 'dark archives' [5], [6], [6], [7].

Archives, therefore, have turned to AI to aid them in the automation of sensitivity reviews [4]. Although helpful, the use of AI in such reviews still has a long way to go before it can be completely relied upon to identify and extract sensitive content. Many issues relating both to the limitations of current AI models and the contextual nature of data privacy remain [4].

A further concern, ironically, is that current data practices for the use of AI to conduct sensitivity reviews risk the very leakage of the sensitive data that archives seek to protect through reliance on commercial or non-privacy preserving AI models [4]. To address this risk and equip archives with capabilities that both protect sensitive data while, at the same time, enabling greater access to archival holdings, we present the first iteration of an architecture and design for a decentralized Web3 digital solution for privacy-preserving AI-enabled archival reference and access. Uysal [8] explains Web3 in the context of transformation that the Internet has undergone over the past few decades. He observes that the Internet began with a "basic and unchanging Web 1.0", proceeded to the more "dynamic and user-centered Web 2.0", and moved to the era of Web 3.0. This latest era is marked by two distinct trajectories of development: the semantic web (which Uysal calls Web 3.0), focused on "organizing and connecting data to improve its accessibility for both humans and machines" and the decentralized web (what has come to be known as Web3, according to Uysal, in order to disambiguate it from the semantic web), focused on strengthening security measures, giving users more control over their data, and promoting transparency, often using blockchain technology [8]. It is Web3 that this paper focuses on.

The remainder of this article is structured as follows: Section 2 presents background and related work. Section 3 presents the Clio-X solution architecture and design in detail. Section 4 reports on a user-based evaluation study that was conducted for assessing user perspectives on the architecture and design of the Clio-X solution, and Section 5 presents a discussion of our findings. Finally, Section 6 discusses planned future work and concludes the paper.

## 2 Background and Related Work

### 2.1 Use of AI in Archives

For this paper, we define AI as a "sub-field of computer science research and practice that aims to develop computational models to simulate or approximate human cognition, behaviour, decision-making, and reasoning' [9] and AI system as "An application that utilizes AI methods, potentially multiple methods, to perform a set of tasks" [9]. Our definition of AI, includes machine-learning, which is an "AI methodology that aims to create a model that can make predictions based on input features." [9].

Within archives and other cultural heritage institutions, the use of AI and AI systems has been growing and, if not yet widespread, is on the rise across all archival work functions [1], [2]. For example, AI is being used to digitize, classify, and restore historical artifacts, as seen in applications that use generative adversarial networks (GANs) for reconstructing damaged artworks [10] and advanced machine learning models for recognizing traditional cultural practices such as weaving or dance [11], [12]. Institutions are leveraging AI to automate metadata creation, improve searchability across large archival datasets [13], and personalize user experiences through adaptive platforms [14]. Projects like Open Dance Lab [12] and Polariscope [15] exemplify how AI can support participatory, community-driven storytelling and cultural knowledge sharing by, respectively, using AI to classify and preserve 3D models of traditional Thai dance movements and in the visualization of digital images, videos, oral testimonies and other documents relating to the territory of Aveiro, Portugal. While AI-enabled archival practice promises to reduce the burden of archival work, a burden arguably impossible to address without such tools [7], there is growing concern about the data practices associated with the use of AI [2], [16]. Some of the issues surrounding AI data practices include: the reinforcement of bias and discrimination in models, potential copyright violations, and the rise of data monopolies that concentrate power among a few entities [2],[16]. Privacy violations and exploitative data collection practices further erode public trust, especially when systems lack transparency and explainability [4]. Furthermore, the environmental and social consequences of unfair AI deployments raise ethical alarms, particularly in contexts where harm disproportionately affects vulnerable communities [4]. Generative AI also risks replicating structures of cultural expropriation and algorithmic homogenization [16]. These issues are compounded by widespread non-compliance with an increasingly complex landscape of data protection and AI regulations [4]. Such concerns threaten to derail the benefits of AI for archives and other cultural institutions unless addressed.

## 2.2 AI and Sensitivity Reviews

As archivists have a mission both to provide access to records and protect personal information, archivists now must comply with a growing body of laws, such as exemption provisions found in freedom of information laws or provisions of data protection legislation, that have arisen as a response to widening concerns about privacy and confidentiality in the digital age [4]. Interpretation and application of these laws adds complexity to the archivists' task of providing access to archival holdings.

Increasingly, archivists are exploring use of AI to address compliance with data protection and privacy laws by automating the sensitivity reviews undertaken before providing public access to materials that could contain personal information. Typically, this entails using AI to identify sensitive information in order to anonymize or redact it before public release [17], [18], [19]

Efforts to identify, anonymize, or redact sensitive information in archival documents have evolved from early, inefficient techniques using regular expressions to more advanced applications of Natural Language Processing (NLP). Researchers have experimented with text classification to flag documents containing sensitive data, but these methods still rely heavily on human input for training and review [17]. Recent tests using AI approaches—such as Term Frequency -Inverse Document Frequency (TF-IDF) with Support Vector Machines (SVMs), and neural networks like Convolutional Neural Networks (CNNs) and Long Short-Term Memory (LSTMs)—have shown limited accuracy, often misclassifying data and struggling with nuanced privacy risks like the 'mosaic effect', where seemingly innocuous data points reveal sensitive information when combined [20]. Additionally, AI models used in these tasks face the challenge of potentially leaking personal information themselves. Model properties can be exploited in attacks, revealing sensitive personal information used in model training [4].

**2.3 Privacy-Enhancing Technologies and Techniques**

Given the current limitations of AI-enhanced approaches to anonymizing or redacting records containing sensitive (e.g., personal) information, researchers have called for research into the application of techniques that enable privacy-enhancing computation over archival data (e.g., [4]and [21]. Privacy-enhancing technologies (PETs) are foundational to the vision of Web3, which emphasizes decentralization, user sovereignty, and secure data sharing without reliance on centralized intermediaries. Tools such as Zero-Knowledge Proofs (ZKP), homomorphic encryption, and Secure Multi-Party Computation (SMPC) enable users to authenticate, transact, and collaborate across blockchain-based platforms while minimizing data exposure [4]. These technologies, along with some others like Trusted Execution Environments (TTEs) or ring signatures, empower individuals to retain control of their personal information (i.e., to achieve 'data sovereignty'), aligning with Web3's goal of creating a more trustless, privacy-preserving digital ecosystem where users can interact transparently yet confidentially without the need to give away or provide researchers with direct access to archival documents (i.e., by using visual analytics) [4]and [21].

**2.4 Compute-to-Data Architecture**

Data sovereignty and privacy protection can be enhanced by Compute-to-Data Architectures (CtD) associated with edge computing approaches [22]. With CtD, data remains in decentralized data stores rather than being moved to centralized repositories to facilitate AI-enabled analysis. Additionally, data stays under the custody and control of an originating source (e.g., the records creator or archives) with AI algorithms 'visiting' the data in a specially provisioned and access-controlled compute environment for purposes of analysis. AI analysis of data is temporarily permissioned and, once completed, typically revoked. And even more importantly, data consumers do not obtain a copy of the original raw data, just the results of the computation. Thus, future reuses should go through the same access-controlled CtD mechanism, guaranteeing that data providers remain in control and sovereign.

Blockchain technology and smart contracts are often used to automate resource access control and service authorization in CtD architecture [23], [24]. Blockchains can enable individuals to retain possession of the private keys that control access to and use of their data (sometimes referred to as 'Self-Sovereign' data management, and often associated with the management of identity data, from which the moniker 'Self-Sovereign Identity,' or SSI, is derived) [25] [26]. Since the data are not moved and access is tightly controlled, CtD architectures are now considered to be much better than traditional centralized architectures for protecting sensitive data and avoiding exploitative AI data practices. Self-sovereign data management is central to the European digital strategy and the associated concept of 'data spaces'. The Data Spaces Support Centre glossary defines a data space as "A distributed system defined by a governance framework that enables secure and trustworthy data transactions between participants while supporting trust and data sovereignty. A data space is implemented by one or more infrastructures and enables one or more use cases" [27]. A report by the Finnish funding agency Sitra notes that "While there are different definitions of data spaces, they all share the same basic objective – to facilitate trusted data flows fairly and transparently for the parties involved in data sharing. In data spaces, individuals and organisations, as data rights holders, are in the driver's seat, deciding who can use their data and on what terms. By comparison, in more centralised and traditional data platforms, the decision-making power is in the hands of a few" [28]

### 2.5 Trusted Execution Environments

Data Spaces and TEEs are complementary components in building secure, federated data ecosystems [4]. Data Spaces facilitate controlled data sharing among participants while maintaining data sovereignty, often across sectors or jurisdictions. TEEs, such as Intel SGX [29] or AWS Nitro Enclaves [30], provide a hardware-based secure environment that ensures sensitive data can be processed without being exposed to unauthorized parties—even the system operator. Together, they enable CtD approaches, where data remains in place and only verified algorithms are allowed to access and process it securely, thus enhancing trust, privacy, and compliance within collaborative data infrastructures.

### 2.6 Privacy-Preserving Machine-Learning

Privacy-preserving machine learning (PPML) refers to a set of techniques and methodologies that enable the training and deployment of machine learning models while minimizing the exposure of sensitive or personally identifiable data. PPML ensures that data privacy is maintained throughout the model lifecycle by using methods such as federated learning, differential privacy, homomorphic encryption, and SMPC (see e.g., [31], [32],[33]. These approaches allow insights to be derived from data without revealing the data itself [4], [34].

### 2.7 Distant Reading and Data Visualization

Distant Reading uses Text Mining, Natural Language Processing, and AI to help researchers analyze large archival corpora [35], [36]. Typically, the output of such analyses is a visualization that represents broad patterns that can be gleaned from archival documents, such as the communication patterns between geolocations, public sentiment over time, or topics or themes represented in a corpus of archival text (see, e.g., [37], [38]). Using Distant Reading enables researchers to learn from large archival corpora without having to inspect and analyze each individual document, which, in turn, can shield sensitive information in the underlying records from exposure and eliminate the need for archivists to conduct sensitivity reviews before providing public access to their holdings [21].

## 3 Solution Architecture

### 3.1 Method

Inspired by [6], which called for the use of design thinking to solve the problem of dark archives, to develop Clio-X's solution architecture we combined Research through Design (RtD) methodology with a human-centred design (HCD) and an Agile Software development methodology.

RtD is a practice-based research methodology that generates knowledge through the iterative process of designing, making, and reflecting [39]. Rather than testing predefined hypotheses, RtD explores research questions by creating artifacts -such as prototypes, systems, or experiences -that embody theoretical and practical insights. The process is exploratory and interpretive, emphasizing situated learning and critical reflection. In the case of our study, we sought to create a medium-fidelity design artifact (i.e., one that sits between a low-fidelity basic, sketch-like artifact and a high-fidelity fully polished, production-ready design) that would allow us to explore the research question: Can Privacy Enhancing Technologies and Distant Reading be used to increase access to sensitive archival documents? In this paper, we discuss how we explored this question from the user's perspective using the Clio-X design artifact.

Human-Centred Design (HCD) is a design methodology that places the needs, experiences, and perspectives of people at the core of the design process [40]. It involves understanding users through empathy and research, defining the problem from their point of view, ideating possible solutions, prototyping, and testing iteratively. HCD emphasizes collaboration with stakeholders, contextual understanding, and iterative refinement to ensure solutions are usable, useful, and meaningful. For this reason, it aligns well with an agile software development approach [41].

Agile software development is an iterative and flexible approach to building software that emphasizes collaboration, user feedback, and rapid delivery of functional components [41]. Instead of following a linear plan, the Agile process divides development into short cycles called sprints, during which cross-functional teams design, develop, test, and refine working software. In our case, we held weekly sprints over a period of four months. Agile is increasingly used in human-centered design, making it well-suited for our project [41].

### 3.2 Solution Description and Architecture

Clio-X is a data space demonstrator that offers a novel decentralized AI-enabled reference and access solution for archives and digital humanities researchers. Clio-X was developed because archival institutions have many archival documents of a sensitive nature and growing researcher requests for access and analysis of these documents using AI. As already mentioned, archives struggle to conduct sensitivity reviews using manual or automated techniques, which prevents compliance with AI and data protection regulations and presents barriers to access for researchers. Clio-X seeks to solve this problem by: (1) allowing archives to make archival datasets available to researchers to run a variety of AI-enabled computations (e.g., exploratory data analysis, clustering, topic modelling, and sentiment analysis) and (2) allowing researchers to access one or more archival datasets to run AI algorithms over data in secure compute environments in such a way that the data never leaves the custody and control of the archives but is still able to return aggregated output results to researchers in the form of statistics or data visualizations.

Clio-X builds upon the Pontus-X [42]technology stack, an open-source reference framework and Web3 'ecosystem of ecosystems' aiming to advance data space technologies for the industrial AI and data economy. Pontus-X was built in 2021 by deltaDAO using the Gaia-X Trust Framework [43], Ocean Protocol (OP) Enterprise [44], and blockchain technology (i.e., OASIS privacy-preserving blockchain [45]. Pontus-X aims to drive cross-industry data collaboration and compliance with European regulations (Data Act, Data Governance Act, AI Act, GDPR) in sectors including, but not limited to, aerospace, space, agriculture, manufacturing, industry 4.0, mobility, AI, smart cities, and data-driven businesses, as well as open science" [43].

Oasis is a Layer 1 decentralized blockchain network designed to be uniquely scalable, privacy-first and versatile. The network has two main architectural components: the consensus layer and the ParaTime layer. Oasis follows a systematic approach for unlocking data for AI analysis with the verification of eligibility, proofs of payments, and acceptance of terms and conditions of service level agreements [25].

Pontus-X relies upon an experimental build of Oasis Sapphire, one of the available ParaTimes within the Oasis blockchain [46]. Oasis Sapphire is a confidential Ethereum Virtual Machine (EVM) compatible blockchain that incorporates several privacy-enhancing features. For example, Oasis provides confidentiality through Trusted Execution Environments (TEEs) on its Sapphire ParaTime layer, which enables encrypted network state and confidential smart contracts that protect data during processing. Developers can build EVM-based decentralized applications (dApps) on Sapphire with smart contracts that are confidential or public [43]. Additionally, Oasis Sapphire offers streamlined dApp development; cost effective functionality with 'free views'; interoperability and compatibility with standard Ethereum tooling and wallets (e.g., MetaMask), and support for off-chain computation [46], [?].

The Pontus-X ecosystem also uses an experimental 'Euro-e', a digital euro ERC-20-based stable coin [43], which enables an instant settlement layer to consume data, software, and infrastructure services in Euros rather than the Oasis protocol's native cryptocurrency token (i.e., ROSE).

Additionally, Pontus-X leverages Ocean Protocol (OP). The overall goal of OP is to allow data providers to share their data in a privacy-preserving manner and consumers to obtain value from accessing and processing the data. The data within the OP marketplace cannot be accessed directly if the CtD approach is used [25]. Data tokens that use the Ethereum ERC-20 token standard are utilized to provide access to the data [25]. The data stays on individual data providers' premises and consumers can only perform computations on the data after accessing the data using a data token [25]. The OP decentralized database maintains a record of every transaction. Any organization with access to a data token has compute access to the data. In other words, each data token provides a license to access the data [25]. Figure 1 provides an overview of the solution architecture of Clio-X. Table 1 describes each component, mapped to the archives and cultural heritage use case context embodied in Clio-X, the technical layer at which the component operates, and who provisions the component in the design of the Clio-X solution.

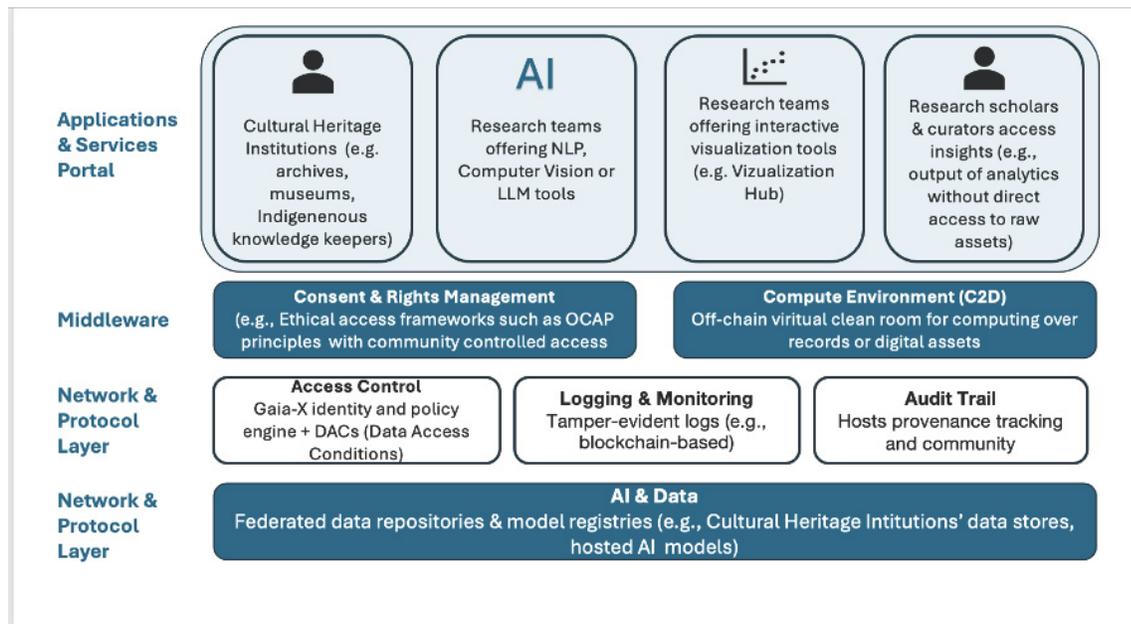

Fig. 1. Overview of Clio-X Solution Architecture (Source: Authors' own adapted from Pontus-X)

Finally, Clio-X offers privacy-preserving machine-learning algorithms that mask sensitive features of archival documents (e.g., names, addresses, social security numbers) prior to performing analytic tasks (e.g., clustering, sentiment classification). Additionally, final results of computations can be imported into a visualization hub for further analysis and exploration of broad patterns and trends without the need to view individual documents that may contain sensitive information.

By leveraging the technical components of Pontus-X and incorporating privacy-preserving machine-learning algorithms coupled with data visualization, Clio-X enables archival institutions to share data, software, and infrastructure services for researchers to consume within a trusted and transparent environment. As envisioned in [4], Clio-X provides a CtD environment, emphasizing privacy-preserving analytics and technical data sovereignty in AI and data spaces. This architecture aligns with principles from the Gaia-X and Ocean Protocol ecosystems, enabling safe and compliant AI computation on private data without data movement.

Table 1. Mapping of Clio-X Architectural Components

| Element | Cultural Heritage Context | Function | Architecture Layer | Provisioner |
|---|---|---|---|---|
| Data Holder | Cultural heritage institutions (e.g., museums, archives, Indigenous knowledge stewards) | Grants permission to use digital records and/or surrogates, restricted records, or metadata. Data holder custom licenses can be specified | Applications and Services | Cultural heritage institutions |
| AI Contributor | Research teams offering NLP, vision, or multilingual models | Trains or applies models for transcription, translation, entity recognition, etc. Clio-X offers privacy-preserving AI models that mask elements of sensitive information in digital records provided by data holders | Applications and Services | Clio-X |
| Visualization Contributor | Research teams offering tools for the visualization of AI output results | Provides tools for visual exploration of AI output results, which permits analysis without direct access to digital records | Applications and Services | Clio-X or Research teams |
| Data Consumer | Research scholars and curators access insights (e.g., output of analytics without direct access to the assets) | Data Consumer: Buys data/compute access. | Application and Services | Researchers |
| Consent and Rights Management | Ethical access frameworks (e.g., OCAP® principles, UNESCO ethics) | Ensures compliance with community-controlled access, usage, and cultural sensitivity. These are articulated in data holder custom licenses | Middleware | Ocean Protocol |
| Compute Environment (CtD) | Secure enclave or virtual clean room infrastructure | Computation occurs without exposing raw records | Middleware | Ocean Protocol for orchestration and Clio-X or cultural heritage institution for clean room |
| Access Control | Gaia-X identity and policy engine + DACs (Data Access Conditions) | Enforces granular access policies e.g., some data might only accessible to researchers with Indigenous community authorship | Network and Protocol Layer | Gaia-X, Ocean Proto-col, OASIS Sapphire blockchain |
| Logging and Monitoring | Tamper-evident logs e.g., blockchain-based | Enables tamper-evident logs of transactions for security and accountability purposes | Network/Protocol Layer | Gaia-X, Ocean Proto-col, OASIS Sapphire blockchain |
| Audit Trail | Hosts provenance tracking and community | Enables provenance tracking and community oversight | Network / Protocol Layer | OASIS Sapphire blockchain |
| AI and Data Stores | Federated data repositories and model registries | Hosts cultural datasets, AI models, and training corpora for exploration | Cloud, Edge, IoT Infrastruc-ture | Data stores e.g., AWS S3 operated by Clio-X or Data Owns e.g., cultural heritage institutions |

### 3.3 Workflow and UX/UI Design

*3.3.1 Foundational Concepts and Components.*

- Clio-X Portal: Web3 portal provided by Clio-X on the Pontus-X marketplace (see, https://portal.pontus-x.eu/)
- MetaMask: Cryptocurrency wallet and browser extension that allows users to manage their EVM-compatible digital assets (e.g., Euro-e) and interact with decentralized applications (dApps) to buy, sell, or gain access to digital assets and services.
- Data NFT: ERC721 token representing copyright/'ownership' of a data service.

- Data Token: ERC20 token used to access compute services over data.
- Metadata: Descriptive fields used to discover data assets; stored with a DID and DID Document (DDO).
- DDO: JSON file with metadata, signatures, and DID information.
- Aquarius (Integrated into the Ocean Node in v4.0 of OP): Metadata cache storing DIDs and DDOs.
- Ocean DB: Manages published data tokens on the marketplace.
- Keeper Node: Nodes that run the Ocean software and make datasets available to the network. Keepers receive newly minted tokens to perform their function. Data Providers need to use one or more Keepers to offer data to the network.
- Compute-to-Data (CtD) environment: Allows AI computations on data without moving it. Runs compute jobs locally at the data provider's site.
- AI Algorithm: AI algorithm used in CtD service.

*3.3.2 Roles in the System.*

- Data Holder: Entity uploading and selling data for Euro-e or offering the data for free.
- AI Contributor: Entity uploading and selling (or offering for free) AI algorithm.
- Provider (Integrated into the Ocean Node in v4.0 of OP): A secure proxy for access control, encryption/decryption, and streamlining access to data or compute services.
- Visualization Contributor: Provides tools for visual exploration of AI output results in a 'Visualization Hub'
- Data Consumer: Buys data/compute access.

*3.3.3 Workflow Summary.*

(1) Data Upload and Publication
    (a) Holder uploads data (e.g., to AWS, Google Drive, or decentralized data store such as IPFS under their control) and, in the Clio-X Portal, uses their MetaMask wallet to publish metadata and data store location in Clio-X catalogue
    (b) Provider encrypts the data location URL and stores it on-chain.
    (c) Metadata and access tokens are published, not the data itself.
    (d) A Data NFT is minted using the Ocean NFT factory, and associated data tokens are created.

(2) Asset Registration
    (a) Metadata, DIDs, and DDOs are stored in Aquarius (now incorporated into the Ocean Node in v4.0 of OP).
    (b) Keeper node registers the asset on-chain (e.g., on the Oasis blockchain) via smart contracts.
    (c) Metadata includes dataset name, description, price, etc.

(3) Discovery and Purchase
    (a) Researchers enter Clio-X Portal and search the Clio-X Catalogue (on the Pontus-X marketplace) using metadata.
    (b) Using their MetaMask wallet, Researcher buys access to data, AI algorithms and compute services with Euro-e (ERC20 Data Token).
    (c) Provider verifies the purchase via the Keeper escrow.

(4) Compute-to-Data Execution
    (a) Provider verifies payment, identity, and consent and rights management acceptance.
    (b) Compute job is submitted and assigned a DID.
    (c) CtD environment is triggered to begin computation.

(5) Job Execution
   (a) CtD environment executes the job in a secure and privacy-preserving compute environment (e.g., a trusted execution environment).
   (b) Execution logs and results are stored in tamper-evident store (e.g., Oasis blockchain).
(6) Post-Compute Access
   (a) Researcher is notified of job completion.
   (b) Researcher can explore job output with visual dashboard, query status, repeat job (before access expires), or try another algorithm.
   (c) Data Holder, AI Contributor, Visualization contributor, CtD environment provider etc. receive payment from escrow upon successful job delivery.

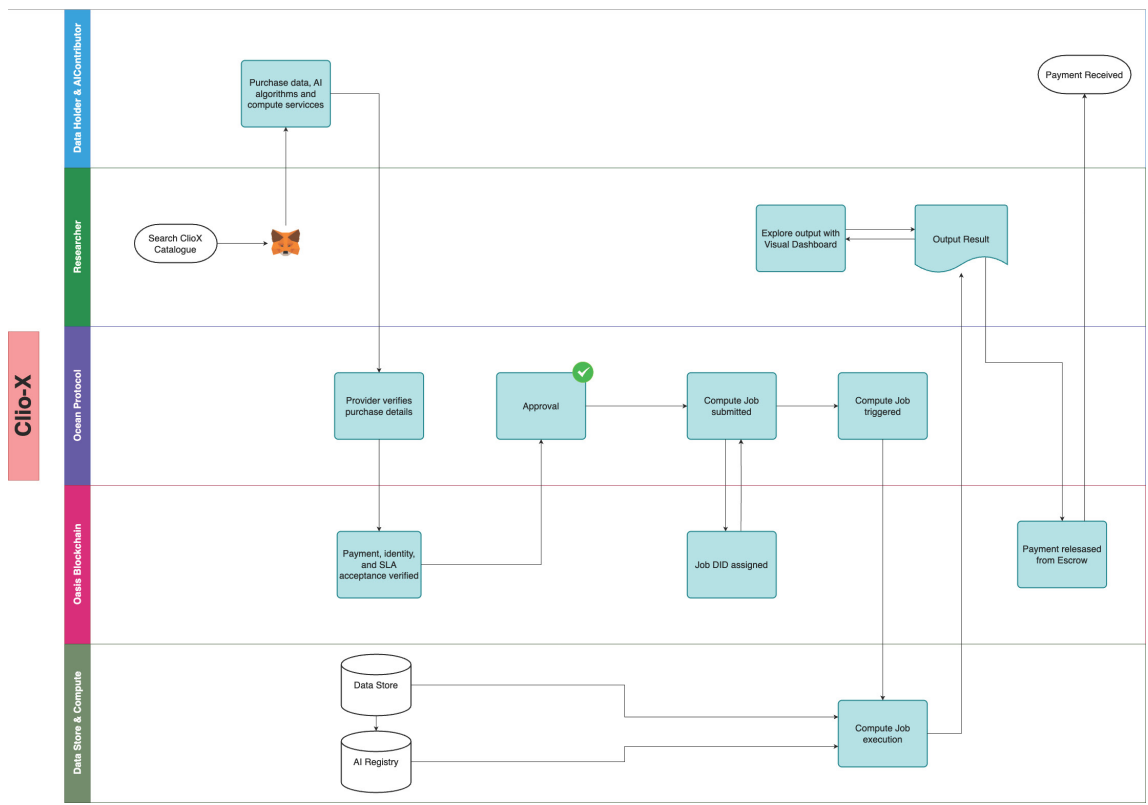

Fig. 2. Overview of Clio-X Workflow (Source: Authors' own)

## 4 User-Based Evaluation

4 *Method.* As mentioned previously, our objective in creating the Clio-X design artifact was to explore from the user's perspective the research question: Can Privacy Enhancing Technologies and Distant Reading be used to increase access to sensitive archival documents? To achieve this objective, inspired by the methodology outlined in [6], we obtained university human-subjects research ethics approval to undertake six separate group or one-on-one demonstration workshops with 14 participants, comprising a mix of archival practitioners and academics from the fields of archival science or library and information science.

Participants were based in Canada, the UK, South Africa, and The United States of America. Recruitment was conducted through invitations sent to contacts within the researchers' professional networks. Interviews took place either on Zoom or in person within a laboratory setting and were recorded and transcribed—either automatically by Zoom or manually by a graduate research assistant.

During the interviews, the researchers presented introductory content to contextualize the study and a medium fidelity prototype of the proposed solution using a dataset of Enron emails [47], a custom, but still rudimentary privacy-preserving AI algorithm and a medium fidelity prototype of a visualization dashboard (see Appendix A for details of an early low fidelity prototype). This was followed by a researcher-led discussion in which study participants were asked a series of questions based loosely upon a semi-structured interview guide (see Appendix B). The resulting transcripts were analyzed using two complementary approaches: thematic analysis conducted by a human coder, and AI-enabled analysis carried out using Atlas.ti (version 25).

Thematic analysis by the human coder followed Braun & Clarke's [48] six-step Reflexive Thematic Analysis framework: familiarization with the data, generating initial codes, searching for themes, reviewing themes, defining themes, and writing up the analysis report. Themes were developed at both semantic and latent levels [49], and ongoing, iterative discussions within the research team enhanced the reflexivity and consistency of the analysis process. From this analysis we manually generated a set of design recommendations (see Table 2)

Analysis with ATLAS.ti entailed conducting a qualitative analysis of the data extracted from the workshops. For this purpose, the files were uploaded into ATLAS.ti, and an initial analysis was conducted using artificial intelligence to identify key concepts (Figure 3). Once this list was generated, a co-occurrence analysis was performed to examine which key concepts in the study were frequently associated with one another (Figure 4). This analysis helped reveal theoretical dependencies between key concepts in our study (e.g., privacy). Additionally, a sentiment analysis was carried out to determine the tone of the conversations, differentiating between three states: positive, negative, and neutral on a sentence-by-sentence basis (Figure 6). The sentiment classification was represented as a Sankey diagram. We then further explored negative associations to uncover underlying attitudes to our platform that could be used to inform our design. Finally, we used the chatbot feature of ATLAS.ti to ask for responses to the questions in our semi-structured interview guide to gather more information about data privacy and then, on the basis of this, to ask for design recommendations (Table 3)

### 4.1 Findings

The two complementary methods of analysis of our user study provided a high-level view of users' perspectives on the design of the Clio-X solution. In the following sections we discuss the results of both our manual thematic analysis and our AI-enabled analysis of the user workshops.

*4.1.1 Thematic Analysis by the Human Coder.*

(1) Trust as Process, Not Product Participants consistently differentiated between the authenticity of the archival data itself—which was largely assumed—and the trustworthiness of Clio-X. In this context, trust was not tied to the datasets, but rather to the platform's verifiability, transparency, and procedural accountability. One participant encapsulated the essence of this recurring theme:

> Instead of being about the records, it's about the process. I would assume that the material itself—the so-called data—is authentic.

As echoed by multiple participants, authenticity was seen as a 'baked-in' baseline feature of the solution. What mattered more was whether users could trace the data and how it was used -how it was handled, curated, and surfaced. This concern aligns with broader archival discourse, where one of the most persistent barriers to using secondary data is inadequate documentation -particularly the lack of information about how data is created and processed. Such information, known as paradata [50],is foundational not only for understanding how to engage with a dataset but often more importantly to assess whether—and to what extent—it can be trusted [51].

(2) Technical Friction, AI Hype, and Professional Precarity While the landscape of the gatekeepers and guarantors of access to trustworthy information for a long time remained relatively stable, since the turn of the millennium, it has evolved rapidly [52], [53], driven by technological developments such as the rapid rise of artificial intelligence. This shift has contributed to an erosion of social trust, a concern voiced repeatedly by participants in relation to the growth of AI and its application within Clio X. A participant expressed skepticism toward the expansion of AI tools, largely due to the perception that their capabilities were being overstated: Generally, when I'm speaking with vendors and they're telling me about this cool AI tool they have they tend to really oversell it. . . they make claims about it that I just know aren't possible with what LLMs are capable of doing. I think there's just a lot of tendency to over promise with these things and then I think there's a desire for people who do hold the purse strings too. . . they're not going to care as much I think as the people who are doing work how effective it is if they see something and they think you know well we have to save $100,000 this year who can we replace with this tool. This skepticism extended beyond individual products to encompass broader cultural dynamics, manifesting in professional networks where trust has become increasingly difficult to establish. Participants described the constant stream of promotional content as overwhelming: Nowadays you see a lot of nice things like this but the paywall just puts you off. Especially now if you look at AI and research—like on LinkedIn—we are drowning right now. Every day you see something new and umm I'm thinking these thought leaders are being paid [chuckles] to promote something. . . they speak about something today and love it, and then two weeks later they say no, move on to this. So, you don't know what to trust, because why should I move on to this? So obviously when you go into that, there's a lot of money to be made, and looking at us right now, our economies are weak. AI was thus not only seen as a tool with technical potential, but as part of a broader cultural and economic landscape marked by marketing fatigue, shifting professional norms, and financial insecurity. In addition, one notable and recurrent theme relating to the technical friction introduced by Clio-X manifested in participants' concern that using the platform might add to their workload, particularly because it demanded expertise that many did not have the time or training to develop: Not many archivists have a lot of time on their hands to be tinkering around with algorithms when you have other work to do.

>It's just about, coming from an information literacy point of view, it's important for people to understand it so they will use it. . . if it it becomes too complicated people tend to shy away and say this is for people who are more clever.

Participants expected technical systems to meet users where they are -both through training and design choices that minimize unnecessary complexity.

(3) Unboxing the Black Box Participants consistently expressed discomfort with the opacity of emerging technologies embedded within the platform, particularly blockchain. The metaphor of the 'black box' was frequently used to describe tools and infrastructures that could not be interrogated by users -ultimately failing to meet their standards of trust or buy-in: So it's a black box. That's my huge problem. It's blockchain. It's a black box. You really don't know anything and a researcher cannot work with a black box. They just need to know what's available. This sentiment was echoed by others who acknowledged their limited understanding of blockchain's potential, especially in relation to their day-to-day archival work. I just have an idea of what blockchain is. I'm kind of like waiting for the day where someone can say 'oh hey this is what it will do in the archives and it's gonna make your life so much easier,' and I know that day will come. Right now, I don't see kind of a use for it in my day to day.

(4) Privacy for Psychological Safety The concerns associated with opaque systems -described by participants as 'black boxes' -also extended to issues of privacy, particularly among those working with sensitive or confidential materials: Since my job primarily relates to privacy I would be really interested in. . . like how well it does at identifying personal information and potentially revealing it... like can it be essentially gamed? Like, could you trick it into giving you more information than it's supposed to? Another participant, grappling with how to express a similar concern, questioned the consistency of the algorithm: Uh, I guess I'm just curious but how can you guarantee the privacy? Like the algorithm you know is taking a person's name and putting 'name' instead but the person's name still exists in. . . I don't know how to pose this question. In both cases, participants focused on privacy not as a policy checkbox, but on the specific mechanisms through which sensitive personal data might be inadvertently disclosed. Their focus was on system design -on whether the platform could anticipate and defend against adversarial interactions, and whether its redaction or anonymization processes could stand up to 'real-world' testing. These reflections underscore a critical distinction: participants were not simply seeking compliance with privacy standards, but an assurance that privacy was built into the architecture of the system. Their trust would only extend as far as the platform's ability to make its handling of risk visible.

(5) Data Visualizations and Analytical Value Word clouds, in particular, were highlighted as a design element that offered visual intuitiveness at the expense of interpretive rigour: One of the biggest problems you have with a dataset like that is you have lexical ambiguity, many words that mean the same thing and many words that have multiple meanings so these tools to my knowledge don't address that . . . 

Participants expressed a desire for analytical affordances: tools that could help distinguish signal from noise, support exploratory analysis, and provide meaningful insights:

> I think given where we are with AI, people are going to expect more powerful analytics. . . I would look into ways to having certain more meaningful like large language models-based analysis of texts.

(6) Governance as Assurance

The governance of the platform -its ownership, institutional affiliations, and funding structure -emerged as a critical determinant of trust. Participants expressed a strong preference for non-profit or open consortium models, which they viewed as more transparent, community-oriented, and ethically grounded than privately held or venture-backed alternatives.

> Consortia works very well. . . it's reassured somebody's not trying to get advantage of you...I saw you open the site and I saw all those companies. . . I don't know if you can fake that but I would like to say that by seeing that it gives some sense of reassurance that I'm working with someone credible.

Trust, in this context, was directly tethered to the governance model. The presence of academic institutions, public archives, or well-known community partners served as a kind of social proof—signalling that the platform was accountable to a collective, rather than driven solely by a single profit motivated entity. For participants, these affiliations were not just symbolic; they suggested a framework of collective responsibility, offering some assurance that if things went wrong, someone could be held to account.

*4.1.2 Design Recommendations.* Based on the above analysis of the workshop transcripts, the human coder derived a set of design recommendations as summarized in Table 2.

4 *Automated Analysis with ATLAS.ti.* As mentioned previously, analysis with ATLAS.ti entailed conducting a qualitative analysis of the data extracted from the workshops. Our first step was to generate a co-occurrence analysis to examine which key concepts in the study frequently associated with one another (Figures 3). This analysis helped reveal theoretical dependencies between key concepts in our study (e.g., privacy, trust, security).

We then further explored word/concept co-occurrence assigning several key concepts a distinct colour to aid our analysis, and focusing on our main theme of data privacy, represented in Figure 4 in a red square denoting its conceptual centrality. The blue squares and the yellow square on the left represent high level abstract concepts in the interviews connected with the theme of Data Privacy: privacy (in general), security and, to a lesser degree, fairness. From this analysis, we observed that privacy and security are completely connected and that workshop participants understood that privacy without security is not possible. As conveyed by the shared yellow colour, the concept of fairness also connected to the themes of communication: copyright and data integrity. The concept of trust is linked to all the other concepts – suggesting that for workshop participants trust was not possible without the other concepts. Several other concepts are represented in the visualization with no colour because they are more specific and concrete (e.g., anonymization). Blockchain is in green because it is a central concept in our solution design (e.g., it is needed to support a privacy-preserving solution architecture).

Additionally, we performed a sentiment analysis of six of the group workshops to determine the tone of conversations in the workshops, differentiating between three states: positive, negative, and neutral on a sentence-by-sentence basis (Figure 5), which is why different classifications may appear within the same paragraph (see Figure 5). While the majority of the texts were classified as neutral, the classifier assigned a negative classification to a high number of the sentences in the workshop transcripts.

Table 2. Design Recommendations Derived from Analysis (Thematic Coding)

| Coding Theme | User Feedback | Design Recommendation |
| --- | --- | --- |
| Trust as Process, Not Product | I would want it to be as reliable as an archivist with five years of experience would be… like if I asked one of my colleagues hey can you look at this file and let me know if you think there's anything worrying in here I would want it to sort of be as reliable as that essentially. | Integrate progressive disclosure to reduce cognitive load while still offering depth for those who seek it. Ensure that metadata and contextual information are always accessible, but not overwhelming. |
| Technical Friction, AI Hype, | Employment is the first thing that comes to mind… will the job that I'm working towards, that I envisioned, be done in the same way a year from now? Yeah, I think there's a lot that's tied into employment like putting food on my table for my kids or paying rent so its much less of a worry of a terminator 2 world or anything its more just like will I be able to keep up and pay my bills. | Minimize friction through adaptive interfaces that cater to varying levels of expertise. Use context-sensitive help, embedded guides, and learning aids. Keep primary interfaces simple and intuitive, with more complex features tucked behind expandable menus. |
| Unboxing the Black Box | How can you prove that the person in the institution that put in that data put in what they said they put in. . . yeah. . . I think it's quite complicated and especially if you're relying on like this third party service to provide a window into data hosted by another institution where like it's sort of necessarily kind of anonymous about how the data set is being hosted in, where it is, and perhaps even like who put it up there right. | Combat opacity with explainable interface elements. Use visual storytelling, traceable data flows, and annotated algorithms to show how outputs are derived. Where possible, allow users to "peek behind the curtain" via interactive components that reveal the logic, not just the re-sults—reducing perceived system ambiguity and fostering control. |
| Privacy for Psychological Safety | something that I want to be kept safe will be kept safe if I'm not paying attention to it like if I don't have my eyes directly on it I can like go home go to sleep and be like well that's going to be safe when I get there in the morning um that's what security means I suppose. | Support user autonomy by offering clear, opt-in settings. Highlight defaults and changes to privacy settings using visual cues that avoid overwhelming users. |
| Data Visualizations & Analytical Value | I never thought that the word trends is particularly meaningful to the sort of average use but I understand that digital humanists might find them useful so I would say that. I think that you know more explanation would be good at what these tools and like use cases. | Ensure that every visualization has an accompanying explanation, legend, or hover-based contextualization to maintain interpretability. |
| Governance as Assurance | Because of my personal and professional values I would say like a consortium or open source um would make me more likely to adopt it. | Display governance clearly in the UI using organizational logos, governance maps, and contactable authority lines. Use design conventions for trust signals—badges, verified partners, or clear affiliations—to reduce uncertainty about who is behind the platform. |

To dig deeper into these results, we used the chatbot feature of ATLAS.ti to explore negative associations connected to underlying attitudes to our solution so that these insights could be used to improve our design (Figure 6). We observed that negative sentiments tended to be linked with confusion, uncertainty and lack of solution transparency and user experience (e.g., expressions such as "I'm still struggling to see"). As a result of these factors, users perceived technical difficulties and lacked trust in the solution.

Finally, we used the chatbot feature of ATLAS.ti to ask for responses to the questions in our semi-structured interview guide (see Appendix B) to gather more information about data privacy and received the following responses:

(1) Anonymization and Synthetic Data: Anonymization is emphasized as a critical factor in preserving data privacy, especially when dealing with sensitive data. Researchers expressed the need for clear pipelines for anonymizing data to facilitate sharing while maintaining privacy (2 voices) The creation of synthetic data is suggested as a method to share data without revealing identifiable information. This approach aims to improve the replicability of research while ensuring that sensitive data remains protected (2 voices) 

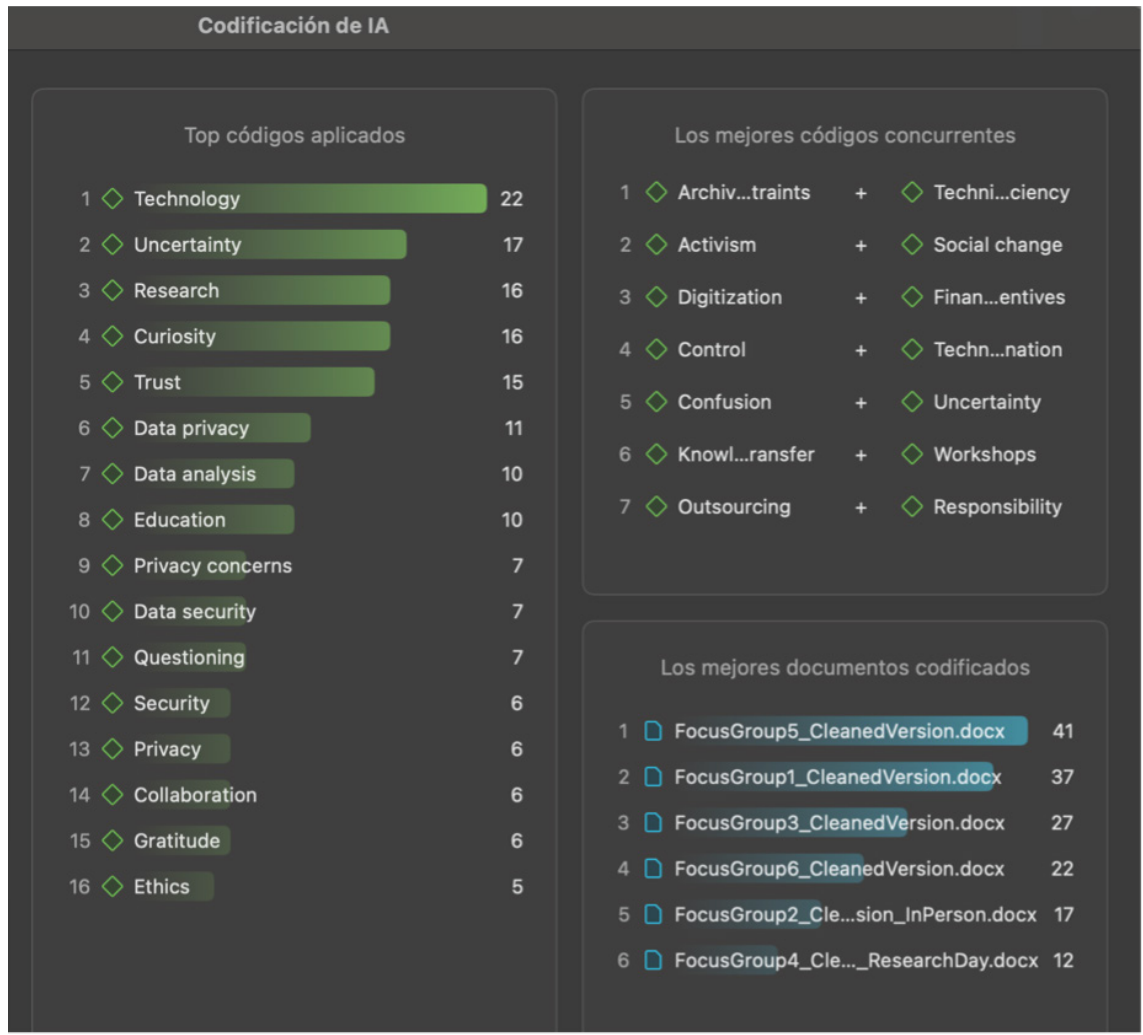

Fig. 3. Codes generated by the AI codification and main co-occurring codes (Source: Authors' own visualization generated using ATLAS.ti)

(2) Control and Custody of Data: A significant aspect of the discussed platform is that researchers do not upload their data directly. Instead, they maintain control over their data in a trusted execution environment, where algorithms can process the data without it leaving the researcher's custody (2 voices). This structure aims to build trust and ensure that data privacy is preserved.

(3) Transparency and Trust: Transparency regarding how data privacy and security are maintained is crucial. Researchers expressed a desire for clear explanations about data storage, processing, and the measures taken to protect privacy (2 voices). Furthermore, trust in the algorithms used is also vital, with discussions around the vetting of algorithms to ensure they do not contain malicious code (2 voices)

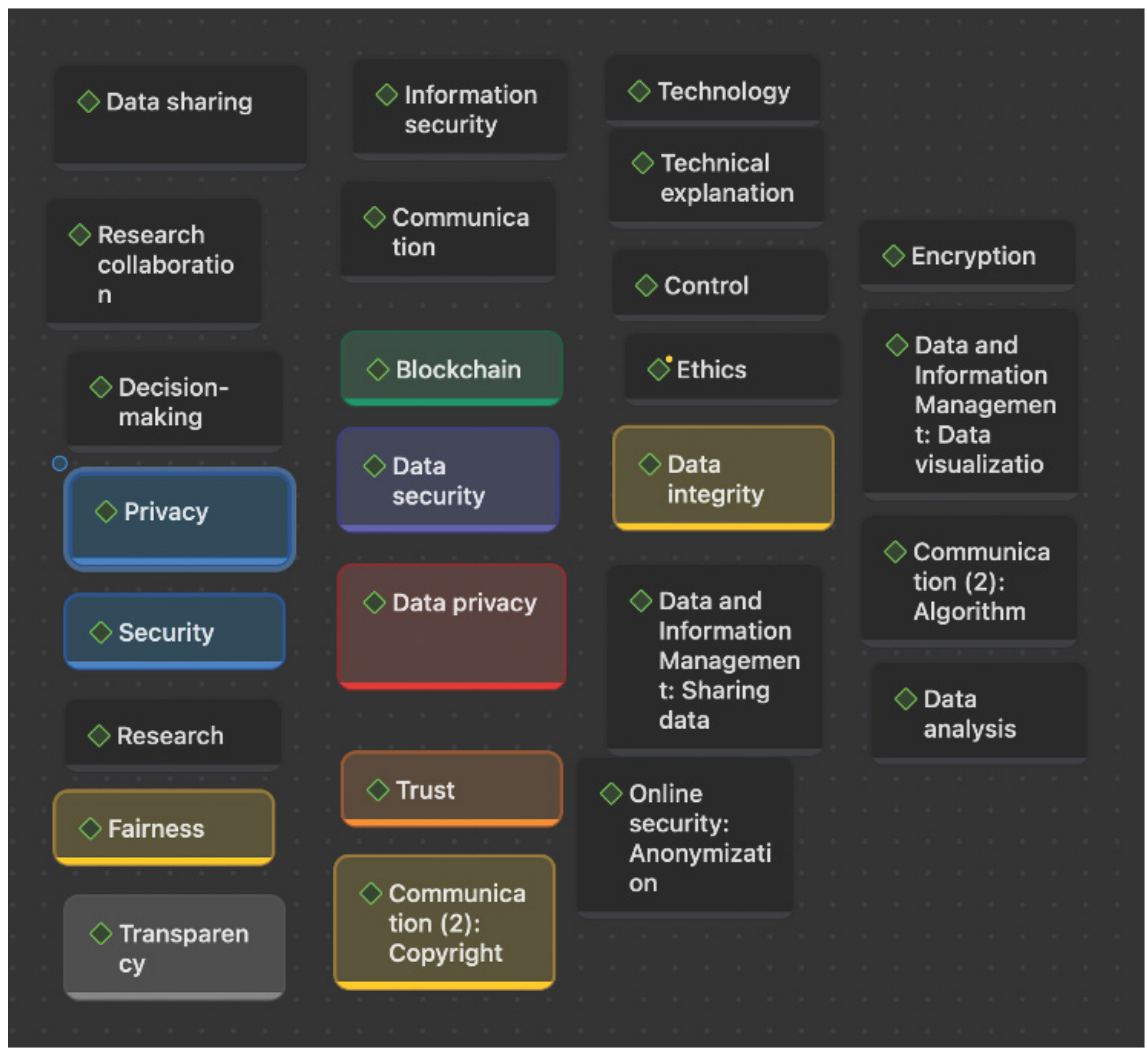

Fig. 4. Data privacy concept and its co-occurring codes (Source: Authors' own visualizaiton generated using ATLAS.ti)

(4) Privacy Impact Assessments: Conducting privacy impact assessments is highlighted as an essential step for institutions adopting new technologies. This process helps identify potential risks and ensures that privacy concerns are addressed before implementation (2 Voices)
(5) Cybersecurity Considerations: Cybersecurity is a growing concern, with discussions about the need for robust measures to protect data as it moves between systems. Ensuring that data remains secure during processing and access is a priority(4 voices)
(6) Community and Consortium Approaches: There is a preference for consortium or open-source models over proprietary solutions, as these are perceived to offer greater transparency and community trust (2 voices). This approach could help mitigate concerns about data privacy and security.

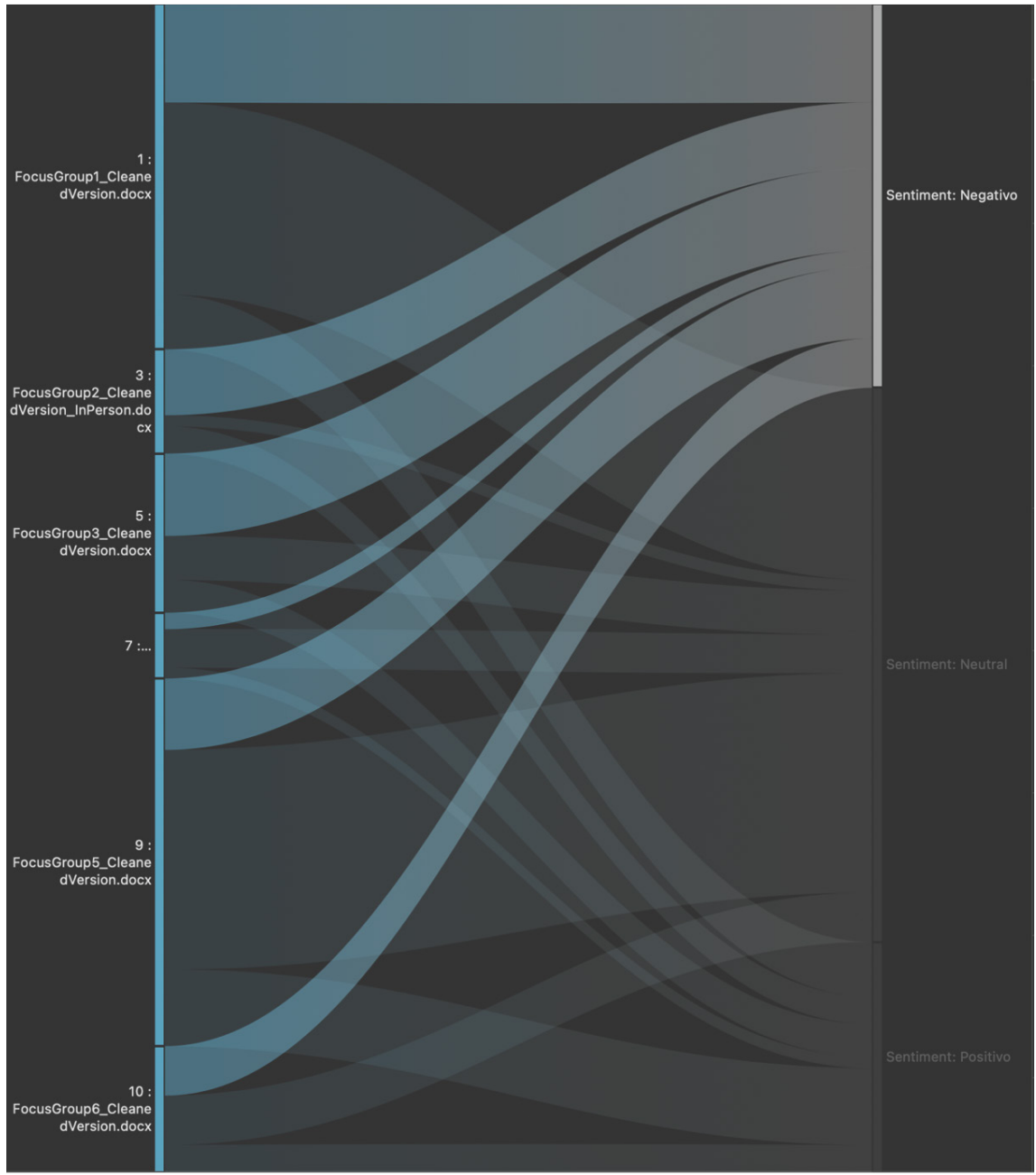

Fig. 5. Sentiment analysis represented as a Sankey Diagram and showing positive, negative, neutral sentiment (Source; Authors' own generated using ATLAS.ti)

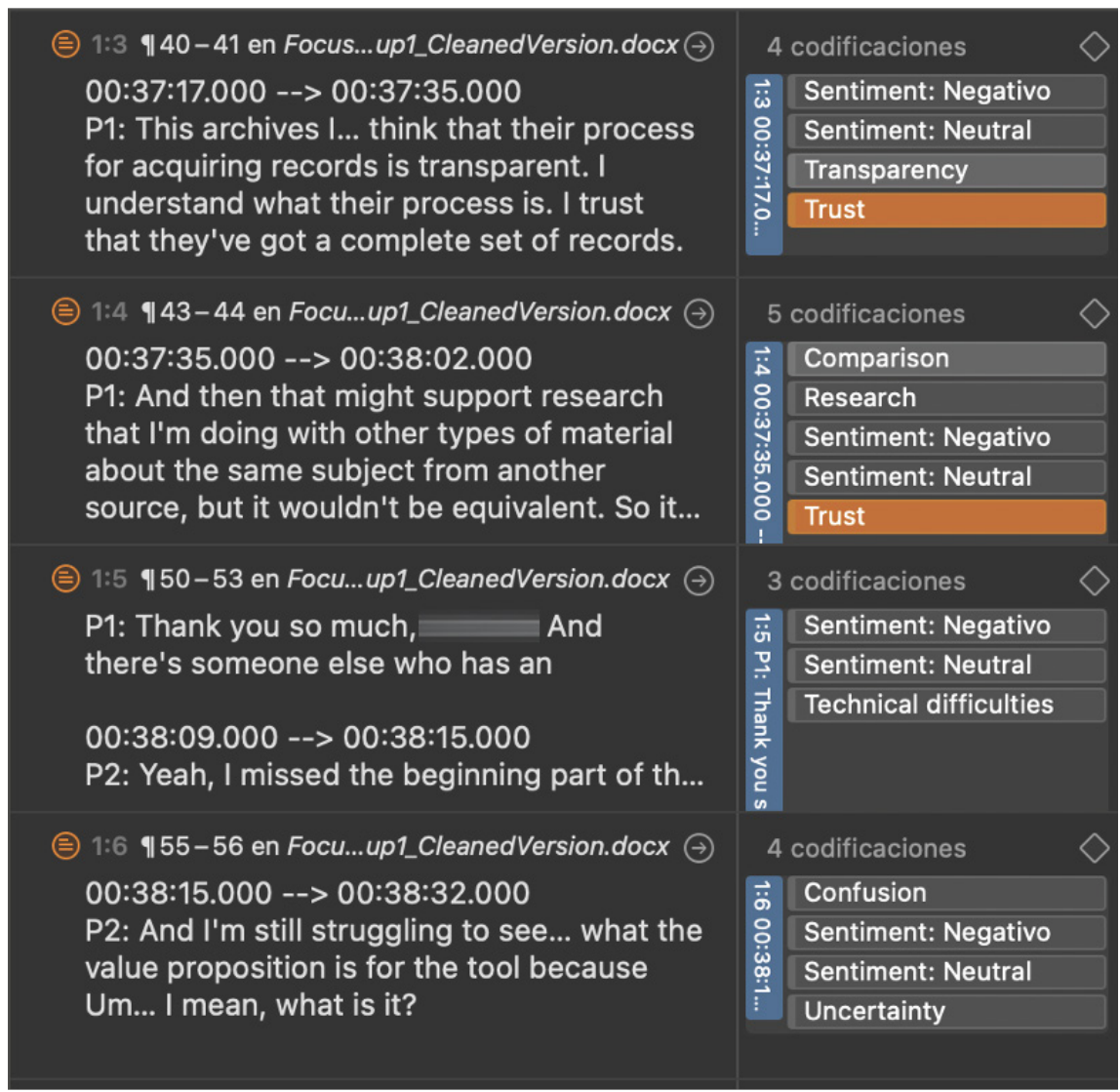

Fig. 6. Voices extracted from the 6 documents and the codes, included positive, negative, and neutral from the sentiment analysis (Source: Authors' own visualization generated using ATLAS.ti)

*4.4.4 Design Recommendations. Based on analysis of the workshop transcripts using ATLAS.ti.* On the basis of this, we asked for design recommendations. An example of results is presented in Table 3, showing the AI prompts we used, the AI design recommendation, and the voices in the workshop transcripts used to generate the design recommendations. The complete list of recommendations follows. 

Table 3. Example Output of Architecture & Design Recommendations Derived from Analysis (Atlas.ti))

| Chatbot Input | Design Recommendations | Voices (some examples) |
| --- | --- | --- |
| Can you provide design recommendations | Ensure that the platform is intuitive and easy to navigate, especially for users who may not have a technical background. This includes clear labelling of features and straightforward workflows (4 voices) | 00:38:32.000 –> 00:38:48.000 P2: So I saw NFT somewhere and I don't know whether that's part of the you know the business case for this tool is this Did I miss something when you were doing the presentation. |

(1) User-Friendly Interface: Ensure that the platform is intuitive and easy to navigate, especially for users who may not have a technical background. This includes clear labelling of features and straightforward workflows (4 voices).
(2) Provide Contextual Help: Implement pop-up explanations or tooltips that provide additional context for features and outputs. This can help users understand complex data visualizations and algorithms without feeling overwhelmed (2 voices).
(3) Scaffolding for Learning: Offer resources that gradually introduce users to technical concepts. This could include tutorials, videos, and hands-on workshops that cater to different learning styles and levels of expertise (4 voices).
(4) Transparency in Algorithms: Clearly explain how algorithms work and what data is being processed. Users should have access to information about the algorithms used, including their purpose and limitations, to build trust in the platform (4 voices).
(5) Customization Options: Allow users to customize their experience by selecting features or parameters they want to adjust. This flexibility can empower users to explore data in ways that are meaningful to them (2 voices).
(6) Training and Support: Provide comprehensive training sessions, both online and in-person, to help users become familiar with the platform. This could include workshops funded by institutions to ensure accessibility (4 voices).
(7) Focus on Data Privacy: Clearly communicate the measures taken to protect data privacy and security. Users should understand how their data is stored, processed, and anonymized (4 voices).
(8) Engagement with Users: Regularly gather feedback from users to refine the platform. This can be done through focus groups or surveys to understand their needs and challenges better (4 voices).
(9) Visualizations and Analytics: Incorporate user-friendly data visualization tools that allow users to easily interpret results. This can include interactive graphs, charts, and dashboards that present data in an accessible manner (2 voices).
(10) Community Building: Foster a sense of community among users by creating forums or discussion groups where they can share experiences, ask questions, and collaborate on projects (2 voices).

### 4.2 Limitations

Our user study consists of only a very small number of workshops and participants. It is possible, therefore, that a larger user study might identify novel insights, or even contradict those we have drawn from this user study. Additionally, our attempts to recruit a greater number of participants from the digital humanities representing the Data Consumer role were largely unsuccessful, and thus we have fewer insights relating to researchers' perspectives on our solution than we would have liked. Nevertheless, the participants in our user study were diverse in geographic location, gender, and professional role, and we achieved saturation (i.e., "the point in coding when you find that no new codes occur in the data. There are mounting instances of the same codes, but no new ones", and "additional data do not lead to any new emergent themes: [54]) in relation to responses to several of our questions. We also note that it was difficult for us 

to cover all the aspects of our solution design and all the questions in our interview guide (see Appendix B), and for some respondents to provide full responses to our questions during the hour-long workshop sessions. A planned future user evaluation of the solution in a real setting will address this limitation.

## 5 Discussion

The emergence of Web3 -marked by decentralization, increased user autonomy, and the integration of blockchain technologies -offers a new digital ecosystem that is both "promising and complex" [8]. On the one hand, it addresses concerns related to centralization, data governance, and, as in the focus of this study, user privacy [8]. However, the technical advances associated with Web3 technologies challenge traditional models of archives management and digital interaction while introducing steep learning curves and conceptual ambiguity for archivists. Given the novelty of Web3, Rogers' 'Diffusion of Innovations' theory [55] offers a useful framework for discussion of our research findings. Rogers describes diffusion as the process by which an innovation is communicated over time among members of a social system. Rogers observed adoption of an innovation depends on how a social system perceives attributes such as:

- Relative advantage: Is it better than what it replaces?
- Compatibility: Does it fit with existing values and practices?
- Complexity: Is it easy to understand and use?
- Trialability: Can it be experimented with on a limited basis?
- Observability: Are the results visible to others?

Communication channels also determine the extent and rate of diffusion: Mass media can raise awareness, but interpersonal networks are often more influential in shaping decisions according to Rogers. Additionally, Rogers saw adoption occuring in five – not necessarily linear, but more iterative, phases: Knowledge (when the user is first exposed to the existence of the innovation and begins to understand how it functions); Persuasion (when the user forms a favorable or unfavorable attitude toward the innovation); Decision (when the user engages in activities that lead to a choice to adopt or reject the innovation); Implementation (when the user puts the technology to use, solving practical and technical issues; and Confirmation (when the user seeks reinforcement for the adoption decision and may reverse it if exposed to conflicting information) [55].

Our two complementary analyses of the user workshops to evaluate the Clio-X solution architecture and design exemplify many aspects of Rogers' Diffusion of Innovations theory by illustrating how potential users of Clio-X navigate the five stages of innovation adoption, with particular emphasis on the role of trust, complexity, social context, and governance structures.

In terms of Knowledge, participants were aware of Clio-X and its features, including AI and blockchain capabilities. However, their understanding was limited, especially regarding how these technologies function (e.g., the metaphor of the 'black box' was repeatedly invoked). There was a desire for clearer, more accessible information, highlighting the importance of well-designed communication channels in Rogers' model.

Aligned with Rogers' understanding of technologies as being spread through cultures and social systems, our study participants' understanding also could have been complicated by what Sloman and Fernbach [56] describe as the "knowledge illusion": the mistaken belief that we understand more than we actually do, due to our reliance on distributed knowledge. They argue the knowledge illusion occurs because we live in a community of knowledge and we fail to distinguish the knowledge that is in our heads from the knowledge outside of it and, therefore, are largely unaware of how little we understand. In the context of our study, social group impressions of Web3, in particular blockchain technology (see, e.g., [57], may well have shaped attitudes and amplified hesitancy to trust Clio-X. The Persuasion stage of Rogers' Theory is most vividly illustrated in study participants' attitudes toward the innovation, which were shaped not just by technical features, but by perceptions of trustworthiness, transparency, and professional values. Participants also distinguished between data authenticity and platform trust, focusing more on process of accessibility and provenance than on the content of data. Concerns were raised about AI hype, opaque algorithms, and privacy risks, feeding skepticism and shaping negative perceptions and illustrating that compatibility with users' values is a critical determinant of adoption.

While the technical capabilities of a Web3 enabled solution like Clio-X represent technological advances, its adoption faces resistance due to how it is perceived and contextualized by potential users. We found that participants were hesitating or rejecting adoption due to technical friction (e.g., usability issues, algorithmic opacity); perceived overcomplexity (especially for time-constrained archivists), and governance concerns (e.g., preference for public or non-profit affiliation as a sign of legitimacy). This reflects Rogers' concept that adoption is influenced by perceived complexity and trialability -here, the inability to meaningfully test or understand the system contributed to resistance. We believe that this limitation will be addressed in a future planned real-world user evaluation of the solution. Despite hesitatancy to adopt, participants did express what would be required to reach adoption. For instance, Clio-X tooling must be more intuitive, offer meaningful analytics, and ensure privacy and safety by design. Additionally, training and system design must "meet users where they are," aligning with Rogers' idea that adoption requires systems to be user-centered. Participants also imagined what would sustain adoption and their trust post-adoption. Factors such as ongoing transparency, strong governance, and social proof from credible institutions would reinforce their decisions to adopt. We also note that the findings pointed to users' trust as not being static -it must be continually validated through visible, accountable practices, aligning with Rogers' notion that users seek reassurance and validation over time.

Our user evaluation of Clio-X richly demonstrates how innovation diffusion is a social, cognitive, and institutional process, not merely a technical one. Participants remain stuck between persuasion and decision stages, with trust as the key mediating factor. For Clio-X to succeed, it must design for transparency, usability, and participatory governance, reducing friction across all stages of adoption. While our user evaluation aligns in many ways to Rogers' Diffusion of Innovation theory, it also diverges from the theory by emphasizing trust as an ongoing process rather than a discrete adoption decision. Participants were less concerned with the innovation's features and more focused on procedural transparency and accountability. Rather than moving through a linear decision-making process, trust was framed as something negotiated over time, especially in relation to systems that handle sensitive data. This challenges Rogers' assumption that innovations are adopted after a clear evaluation of their advantages. Another key departure lies in the erosion of social influence as a driver of adoption. While Rogers highlights the role of opinion leaders in persuading early adopters, participants in this case expressed skepticism toward promotional discourse, influencers, and constantly shifting trends. Rather than fostering adoption, these signals generated distrust and fatigue, undermining the credibility of those promoting the platform. The case also highlights the limits of Rogers' assumptions around complexity, trialability, and observability. Participants described Clio-X as a 'black box' -technically opaque and difficult to evaluate. They lacked the time, resources, or training to experiment with or meaningfully observe its capabilities. These barriers were not easily overcome by better communication or training, suggesting a more intractable form of technological alienation than Rogers' model anticipates. Further, the theory underestimates the influence of structural and economic factors. Participants were wary of the solution not only due to technical concerns but also because of broader anxieties about automation, professional precarity, and institutional funding cuts. These concerns point to the importance of political economy in shaping innovation resistance -factors that Rogers' theory treats as background rather than central. Finally, governance emerged as a key condition of adoption, not a secondary consideration. Participants emphasized the need for transparent, community-based oversight structures -such as public consortia or academic partnerships -before they could trust the platform. This emphasis on institutional ethics and collective accountability stands in contrast to Rogers' more individualistic framing of the adoption process, albeit that adoption decisions are influenced by culture and the social system -revealing a gap in how the theory accounts for the socio-political aspects of trust in digital systems.

## 6 Conclusion and Future Work

In an era of big data, archivists face the dual challenge of managing vast volumes of digital records, many containing sensitive personal or cultural content, and complying with increasingly stringent regulatory requirements that demand careful review of this material. While artificial intelligence appears to be the only scalable means to meet this challenge, current AI technologies fall short, particularly in ensuring transparency, interpretability, and privacy needed in archival contexts.

Clio-X, a Web3-based solution, was designed to address these gaps by preserving archival control over data and embedding privacy-enhancing technologies into its architecture. Technically, Web3 offers the potential to affirmatively answer yes to our core research question (i.e., "Can Privacy Enhancing Technologies and Distant Reading be used to increase access to sensitive archival documents?").

However, a user evaluation of a medium-fidelity Clio-X prototype revealed a more complex picture. Drawing on Rogers' Diffusion of Innovations theory, the findings indicate that while users conceptually welcomed the solution, actual adoption was hindered by opaque system design as well as concerns about economic precarity, AI hype, and governance. Trust was framed not as a static outcome of the adoption of the solution but as a dynamic, process-oriented requirement, particularly for systems handling sensitive data. As such, while Clio-X may be technically well-aligned with archival needs, meaningful adoption will require addressing deep-seated concerns around usability, legitimacy, and ethical accountability.

Our future work will aim to address these outstanding issues through implementing the design recommendations derived from our study findings, engaging with critical perspectives on AI-adoption in archives to support ethical accountability, and leveraging distinct Web3 forms of decentralized governance, in particular, the Decentralized Autonomous Organization (DAO) -a blockchain-based governance structure that operates through transparent, programmable rules encoded in smart contracts with collective decision-making by its members, who typically hold governance tokens that allow them to propose, deliberate, and vote on changes or actions. We believe a Clio-X DAO could enhance user perception of the solution's legitimacy. By embedding participatory governance into its technical infrastructure (a hallmark of Web3 and blockchain) Clio-X has the potential not only to meet the functional demands of ethical AI use in archives, but to earn the trust of its users through sustained transparency, accountability, and community stewardship.

## Acknowledgments

To be added.

**A Sample Screenshots of Presention Shown to Study Participants**

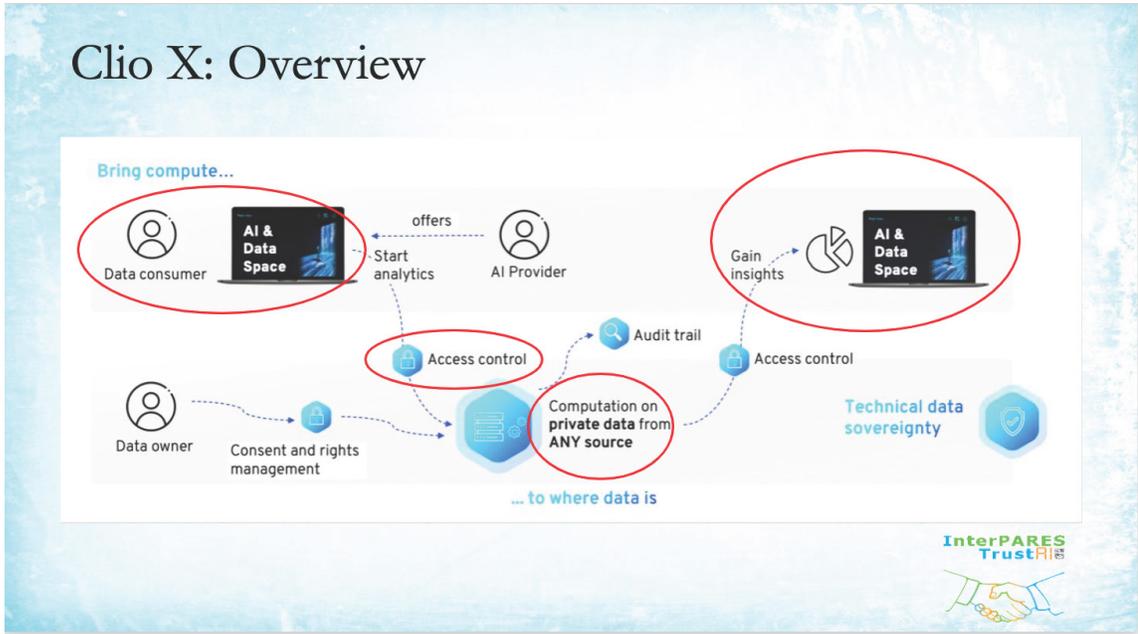

Fig. 7. Screenshot of presentation slide used to explain the solution architecture to study participants

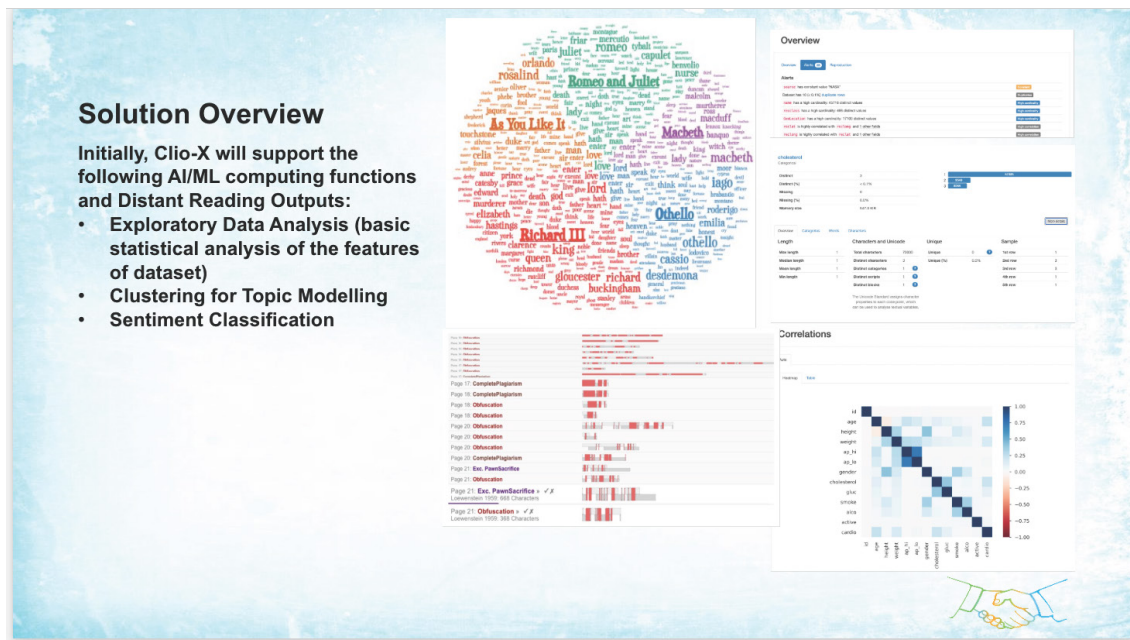

Fig. 8. Screenshot of presentation slide used to explain the solution's distant reading and visualization capabilities

## B Interview Questions

(1) Conversation starter question
    - 
    - What does security mean to you?
(2) Visualizations
    - What are your impressions of the visualizations presented for accessing archival datasets?
    - Did the visualizations feel intuitive, or were there areas where you needed additional context or guidance?
    - What changes or additions to the platform's design would make it more user-friendly for your research needs?
(3) Blockchain and Transactions
    - After seeing the demonstration, what are your thoughts on using blockchain technology for purchasing archival datasets?
    - How comfortable do you feel with the process of using a digital wallet like MetaMask for transactions on this platform?
    - Does the idea of paying with digital euros feel straightforward or raise any concerns for you?
    - Would the use of digital euros affect your likelihood of adopting the platform? Why or why not?
(4) Trustworthiness and Privacy
    - What specific features do you look for in online platforms to assess the trustworthiness of data?
    - Are there specific privacy features in the platform that stand out as particularly useful, or any you feel are missing?
(5) Adoption Factors and Pricing 

- What factors would most influence your decision to adopt this platform for accessing archival datasets? (e.g., energy consumption, cryptocurrency, sentiments on blockchain technology)
- How much would you be willing to pay for access to archival datasets? Would you say €50-€200, €200-€500, or €500-€800 per dataset, depending on content and access level?
- Which management model would make you most likely to adopt this platform: an open-source project, a non-profit organization, a private company, or a consortium of archives?

(6) Data Authenticity and Value
- What would help you feel confident in the authenticity of the archival data provided on this platform?
- How would you assess the value of archival data provided here compared to other sources you've used?

(7) Blockchain Sentiments and Environmental Impact?
- What are your overall sentiments toward blockchain technology in archival work—do you see it as a useful tool, a potential challenge, or both?

(8) Privacy and Trustworthiness
- Are there specific privacy features in the platform that stand out as particularly useful, or any you feel are missing?
- Usability and Adoption Factors
- What changes or additions to the platform's design would make it more trustworthy for archival purposes?
- What changes or additions to the platform's design would make it more user-friendly for archival purposes?
- Are there specific features from other platforms that you think would be beneficial to include here?
- What factors would most influence your decision to adopt this platform for accessing archival datasets? For example, data privacy protections, ease of use, cost, or transparency of data sources?
- Which management model would make you most likely to adopt this platform: an open-source project, a non-profit organization, a private company, or a consortium of archives?